\documentclass[abbrv,aps,prb,twocolumn,showpacs,preprintnumbers,amsmath,amssymb,
nofootinbib,superscriptaddress]{revtex4}

\usepackage{amsmath,amssymb,amsfonts}
\usepackage{bm,graphicx,color}

\begin{document}
\title{Doping induced spin-state transition in LaCoO$_3$: dynamical mean-field study}
\author{P. Augustinsk\'y}
\affiliation{Theoretical Physics III, Center for Electronic Correlations and Magnetism, 
Institute of Physics, University of Augsburg, D-86135, Augsburg, Germany}
\affiliation{Institute of Physics, Academy of Sciences of the Czech republic,
Cukrovarnick\'a 10,
Praha 6, 162 53, Czech Republic}
\author{V.~K\v{r}\'apek}
\affiliation{Institute of Physics, Academy of Sciences of the Czech republic,
Cukrovarnick\'a 10,
Praha 6, 162 53, Czech Republic}
\affiliation{Central European Institute of Technology, Brno University of Technology, Technick\'a 10, 616 00 Brno, Czech Republic}
\author{J.~Kune\v{s}}
\email{kunes@fzu.cz}
\affiliation{Institute of Physics, Academy of Sciences of the Czech republic,
Cukrovarnick\'a 10,
Praha 6, 162 53, Czech Republic}
\date{\today}

\begin{abstract}
The hole and electron doped LaCoO$_3$ is studied using the dynamical mean-field theory. The 
one-particle spectra are analyzed and compared to the available experimental data, in particular
the x-ray absorption spectra. Analyzing the temporal spin-spin correlation functions
we find the atomic intermediate spin state is not important for the observed
Curie-Weiss susceptibility. Contrary to commonly held view about the roles 
played by the $t_{2g}$ and $e_g$ electrons we find narrow quasi-particle bands
of $t_{2g}$ character crossing the Fermi level accompanied by strongly damped $e_g$ excitations.
\end{abstract}
\maketitle

The electron-electron repulsion within the partially filled $d$ shell
is for a long time known to place the transition metal oxides
among the most puzzling materials with properties
varying substantially upon a small change of temperature, pressure,
or carrier concentration. More recently the Hund's intra-atomic exchange
and the competition between various spin states come to the spotlight 
as one of the key features in the physics of iron pnictides.~\cite{yin11,chaloupka12}
However, the spin state competition is a much broader phenomenon~\cite{georges13}
and we are only starting to understand its implications for the properties
of materials.

LaCoO$_3$ is a prototypical system with competing spin states.
A small gap non-magnetic insulator at low temperature
acquires Curie-Weiss (CW) susceptibility above 100~K~\cite{jonker53}
followed by disappearance of the charge gap between 450 and 600~K.~\cite{tokura98}
The generally accepted explanation is essentially atomic physics 
of the low spin ($t_{2g}^6e_g^0$) ground state with a nearby excited
state of either high spin (HS, $t_{2g}^4e_g^2$) or intermediate spin (IS, $t_{2g}^5e_g^1$)
character. The nature of the excited state is still debated.
Hole doping of the Co $d$ bands by substitution of La with Sr, Ca or Ba  
also leads to a strong magnetic response.~\cite{itoh94,kriener04}
At $<18\%$ Sr concentration microscopically inhomogeneous phase is observed that
can be described as magnetic clusters separated by non-magnetic insulating matrix.
Above $18\%$ Sr concentrations
the material becomes a homogeneous ferromagnetic (FM) metal. 
The interplay of spin state competition with the itinerant electron physics
opens interesting possibilities~\cite{chaloupka12}, which are yet to be investigated.

Properties of La$_{1-x}$Sr$_x$CoO$_3$ are commonly discussed in the context of double-exchange
model which provides a satisfactory description of related La$_{1-x}$Sr$_x$MnO$_3$ 
family. In this picture the $t_{2g}$ electrons are localized on the metal atom
forming the local spin moment while $e_g$ electrons form dispersive bands.
Numerous comparative studies, however, found sizeable differences between
cobaltites and manganites. Colossal magnetoresistance, the hallmark of 
manganite physics, is not found in cobaltites.~\cite{paras01,kriener04}
The NMR relaxation rates in cobaltites are several orders of magnitude
larger than in manganites.~\cite{hoch07} The linear specific heat coefficient
in La$_{0.7}$Sr$_{0.3}$CoO$_3$ is 16 times larger 
than in La$_{0.7}$Sr$_{0.3}$MnO$_3$.~\cite{paras01} These observations raise 
the question of the relevance of the double-exchange picture for
doped cobaltites. 

The strong $T$-dependence of physical properties even in the parent compound
LaCoO$_3$ make theoretical description of cobaltites challenging. 
Several density functional and Hartree-Fock calculations on doped cobaltites
were reported.~\cite{abbate94,takahashi98,ravin02,knizek10,medling12} 
In this Letter we use the combination of the dynamical mean-field theory \cite{dmft}
and the density functional theory (DFT+DMFT) \cite{ldadmftb,ldadmftc} to study the one-particle spectra
and magnetic properties of doped LaCoO$_3$. Our results show
an evolution with doping from a stoichiometric insulator, where electronic correlations are hidden
on the one-particle level, to a FM metal, where a sizable many-body
renormalization and quasi-particle (QP) damping take place.
The DMFT solutions in the hole-doped regime are characterized by 
strongly damped $e_g$ excitations and dispersive QP bands of $t_{2g}$ character
crossing the Fermi level thus contradicting the double-exchange picture.
The local moments which appear upon hole doping are found to originate in the Co HS state.

The calculation proceeds in several steps. First, the LDA
band structure is determined using WIEN2k \cite{wien2k} and 
an effective Hamiltonian spanning the 
Co $d$ and O $p$ bands is constructed.~\cite{wannier90, w2w}
Although recently doubts have been raised concerning the reliability
of $p-d$ model for the Mott-Hubbard systems such as V$_2$O$_3$
and LaNiO$_3$~\cite{parragh13}
the strong Co-O hybridization~\cite{krapek12} and 
its role in stabilization of the LS state~\cite{haverkort06}
suggest that the explicit inclusion of O~$p$ states is
important for the physics of LaCoO$_3$.
For all dopings we used the same Hamiltonian obtained for
the 5~K lattice parameters of the stoichiometric LaCoO$_3$. 
Adding the explicit electron-electron interaction within 
the Co $d$ shell we arrive at the multi-band
Hubbard Hamiltonian of the form
\begin{equation}
 H = \sum_\mathbf{k}
\begin{pmatrix}\mathbf{d}^\dagger_\mathbf{k} \\
\mathbf{p}^\dagger_\mathbf{k} \end{pmatrix}
\begin{pmatrix} h^{dd}_\mathbf{k} - E_\text{dc} & h^{dp}_\mathbf{k} \\
h^{pd}_\mathbf{k} & h^{pp}_\mathbf{k} \end{pmatrix}
\begin{pmatrix}\mathbf{d}^{\phantom\dagger}_\mathbf{k} \\
\mathbf{p}^{\phantom\dagger}_\mathbf{k}\end{pmatrix}
+\sum_i W^{dd}_i.
\end{equation}
Here, $\mathbf{d}^{\phantom\dagger}_\mathbf{k}$
($\mathbf{p}^{\phantom\dagger}_\mathbf{k}$)
is an operator-valued vector whose elements
are Fourier transforms of $d_{i\alpha}$ ($p_{i\gamma}$), which annihilate
the Co $d$ (O $p$) electron in the orbital $\alpha$ ($\gamma$) in the $i$th unit cell.
The on-site interaction $W^{dd}_i$ is approximated by the density-density form
with parameters $U$=6~eV and $J$=0.8~eV~\cite{krapek12}(for the full interaction 
matrix see SM). Comparison to the rotationally invariant interaction
for similar materials is discussed in Refs.~\onlinecite{kunes12,krapek12}.
To justify the density-density approximation we point out that the presented 
results hold also in the ferromagnetic phase with broken spin-rotation symmetry
as discussed later.
The double-counting term $E_\text{dc}$ approximately
corrects for the explicitly unknown mean-field part of the interaction
coming from LDA. It was chosen to equal the orbitally averaged Hartree part 
of the self-energy.~\cite{kunes09}
The hybridization expansion continuous-time quantum Monte Carlo \cite{werner06,alps} with the improved
estimator for the self-energy \cite{august13} was used to solve the auxiliary impurity problem.
All the real-frequency spectra were obtained by 
analytic continuation of the self-energy using maximum entropy method.~\cite{maxent}
The calculations were performed at the temperature of 580~K.

\begin{figure}
  \begin{center}
    \includegraphics[height=0.48\columnwidth,angle=270,clip]{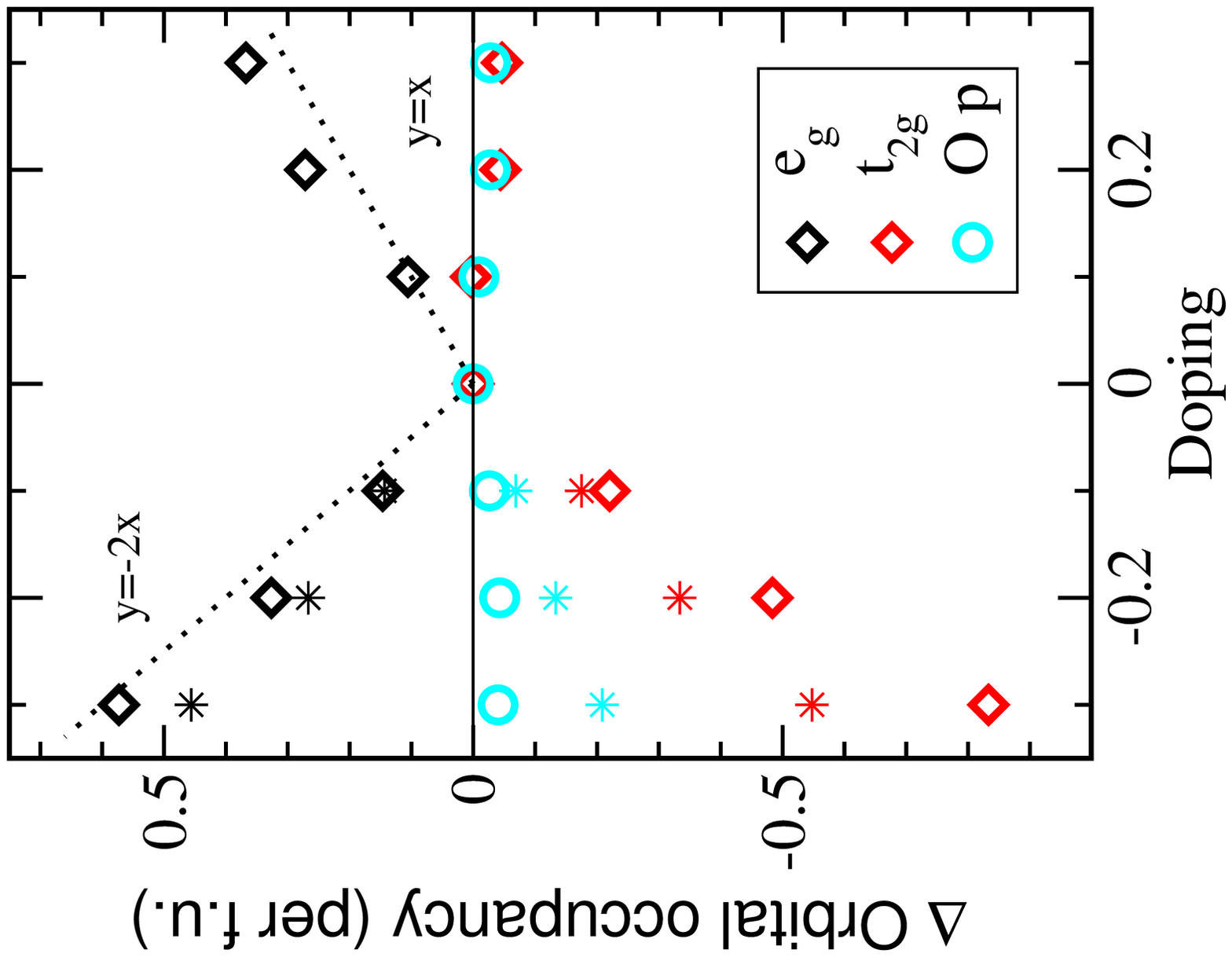}
  \includegraphics[height=0.42\columnwidth,angle=270,clip]{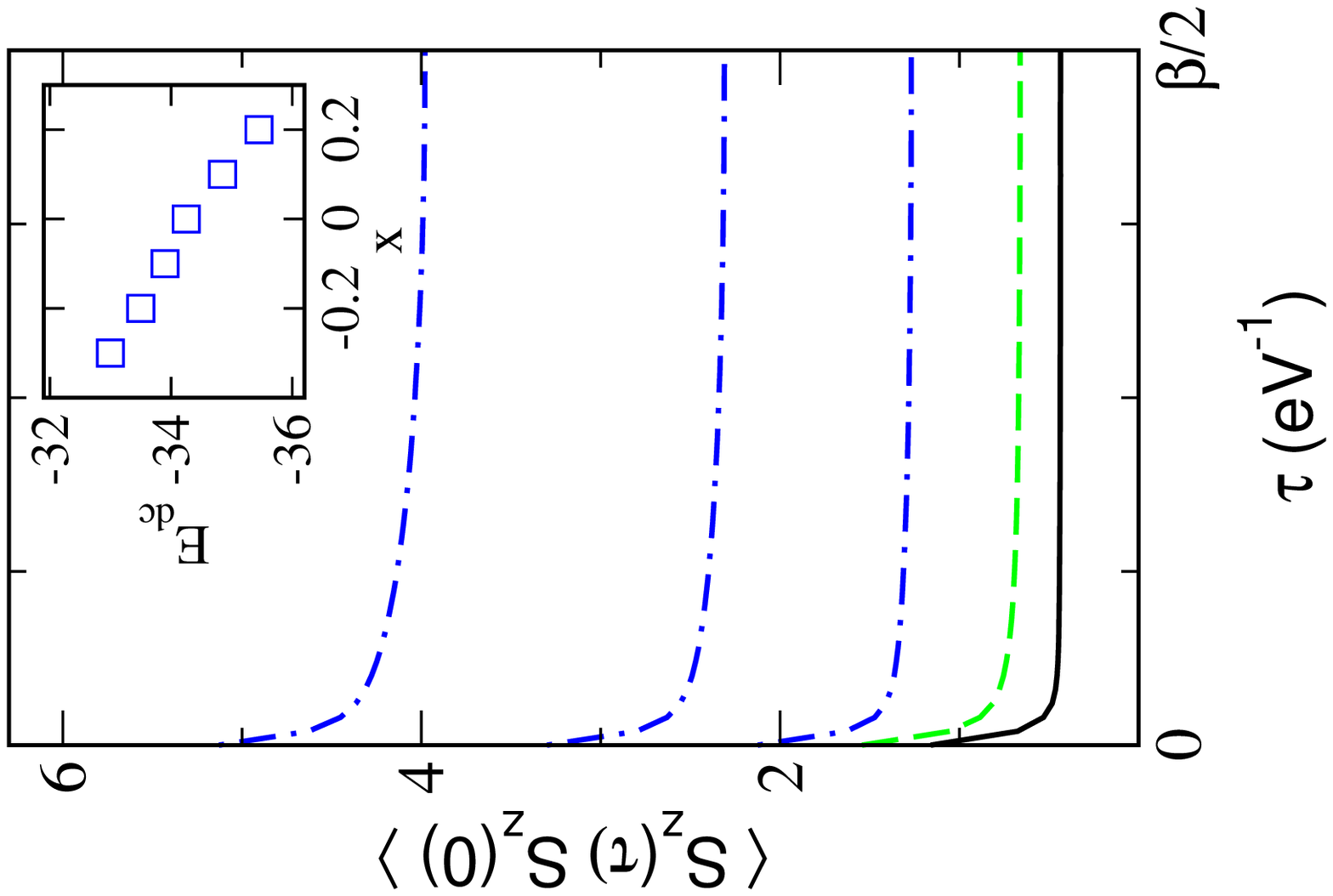}
  \end{center}
\caption{\label{fig:occ} (color online) Left panel: the doping dependence (hole doping
$<0$) of the orbital occupancy. The stars mark results obtained with fixed double-counting correction.
Considering the FM order for x=0.3 changes the data by less than 5\%.
Right panel: the local spin-spin correlation functions: black (full line) -- the undoped system,
blue (dash-dotted line) -- hole doping x=0.1, 0.2, 0.3 (corresponding to increasing magnitude), and
green (dashed line) -- electron doping of 0.2. The inset shows the doping dependence of $E_{\text{dc}}$.}
\end{figure}

The one-particle spectrum of the undoped system~\cite{krapek12}
has a character of an uncorrelated band insulator with a
gap between the filled $t_{2g}$ and empty $e_g$ states.~\cite{note1}
Doping leads to a markedly non-monotonic variation of the orbital occupancies
(Fig.~\ref{fig:occ}) indicating a departure from independent particles behavior.
In particular on the hole-doped side, approximately two $e_g$ electrons
and three $t_{2g}$ holes are created for each hole added to the system.
This is accompanied by appearance of fluctuating local moments 
as reflected by growing spin-spin correlation functions at long times
in the absence of static order, $\langle S_z\rangle=0$
(right panel of Fig.~\ref{fig:occ}). 

Before we analyze the origin of the local moments we establish
that the result is not an artefact of our choice of the double-counting
correction $E_{\text{dc}}$.
In addition to the self-consistently adjusted $E_{\text{dc}}$, 
used throughout the paper (see inset in Fig.~\ref{fig:occ}), 
we have performed test calculations with $E_{\text{dc}}$ fixed to its $x=0$ value. 
Fixing $E_{\text{dc}}$ moves, with increasing hole doping, the $d$ and $p$ bands
towards each other which promotes the LS state. Nevertheless
even with $E_{\text{dc}}$ varied in the unfavorable direction
the observed effect is robust (Fig.~\ref{fig:occ}).
\begin{figure}
  \begin{center}
    \includegraphics[width=0.9\columnwidth,clip]{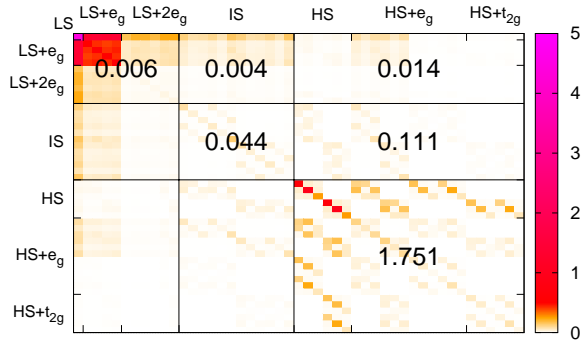}
  \end{center}
\caption{\label{fig:correl} Atomic state correlation matrix
$\Pi_{\alpha\beta}$ for the
hole doping of x=0.2 with the contribution of the most abundant $d^6$, $d^7$ and $d^8$ states (see Ref.~\onlinecite{krapek12} for the notation).
The numbers show the contributions to the local susceptibility $T\chi_{\text{loc}}$ summed
over the blocks.}
\end{figure}

Associating the local moment responsible for CW response of an atom embedded in solid 
with a particular atomic multiplet may not be possible 
in a strongly hybridized and even metallic system.
In Ref.~\onlinecite{krapek12} we have introduced
the atomic state correlation matrix 
$\Pi_{\alpha\beta}=\int_0^{1/T}d\tau\langle P_{\alpha}(\tau)P_{\beta}(0)\rangle$
to address this question in undoped LaCoO$_3$.
$\Pi_{\alpha\beta}$ includes effects of both quantum and statistical fluctuations and
provides information about the likelihood of finding
the Co atom in state $\alpha$ (column sum) as well as about the
average duration of the visit to a given state 
(large diagonal elements mean long visits). 
Weighted with the spin matrix elements, $S^z_{\alpha}S^z_{\beta}\Pi_{\alpha\beta}$
is the contribution of the pair $\alpha \beta$ to the local susceptibility.
The correlation matrix in Fig.~\ref{fig:correl} reveals
that hole doping causes increasing weight of the HS block
which dominates the local susceptibility,
similar to the thermal effect in stoichiometric LaCoO$_3$.~\cite{krapek12,zhang12}
\begin{figure}
  \centering
  \begin{minipage}{0.45\columnwidth}
   \centering
   \includegraphics[height=\columnwidth,angle=270,clip]{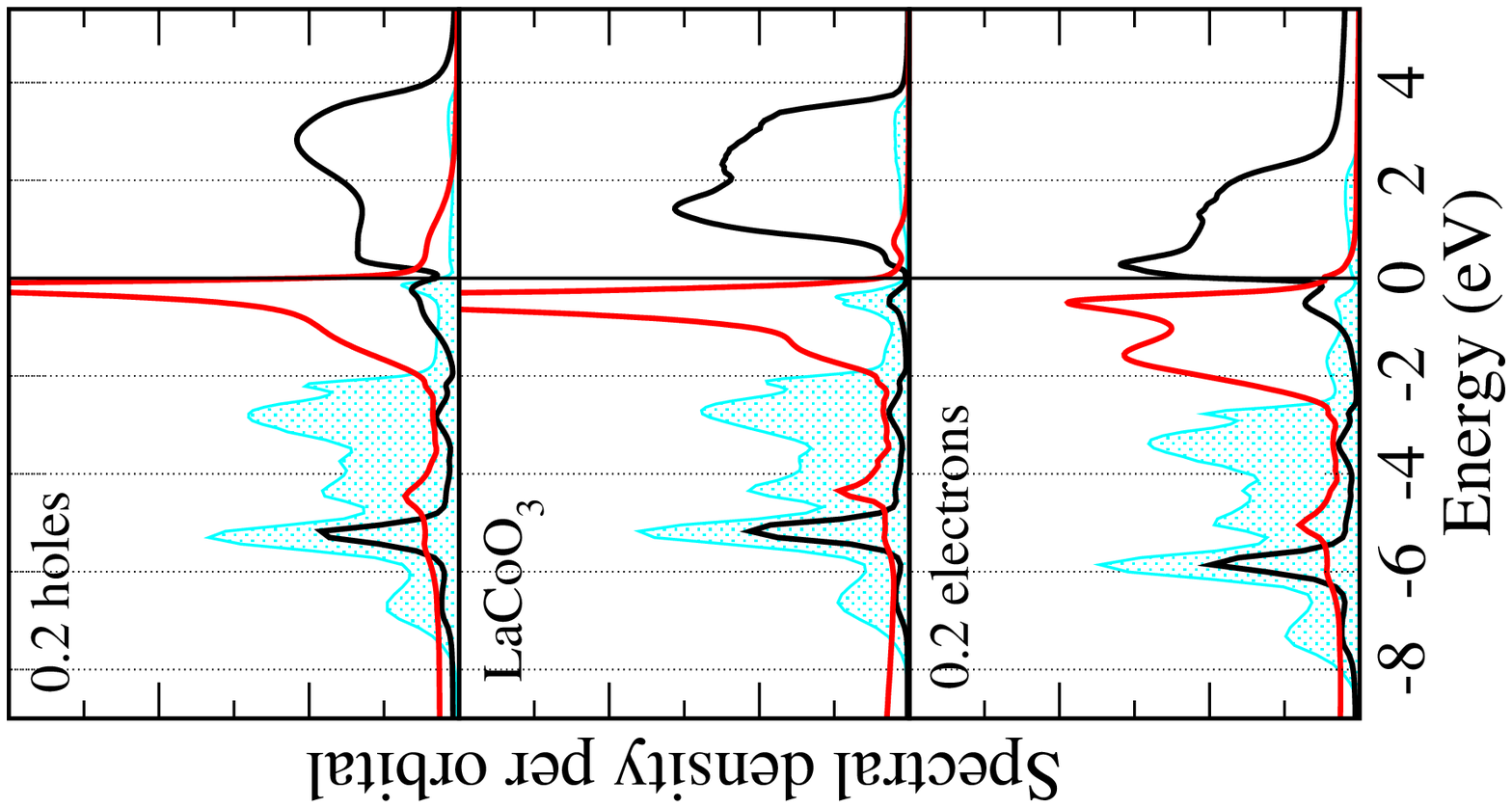}
  \end{minipage}
  \begin{minipage}{0.45\columnwidth}
  \includegraphics[width=0.9\columnwidth,clip]{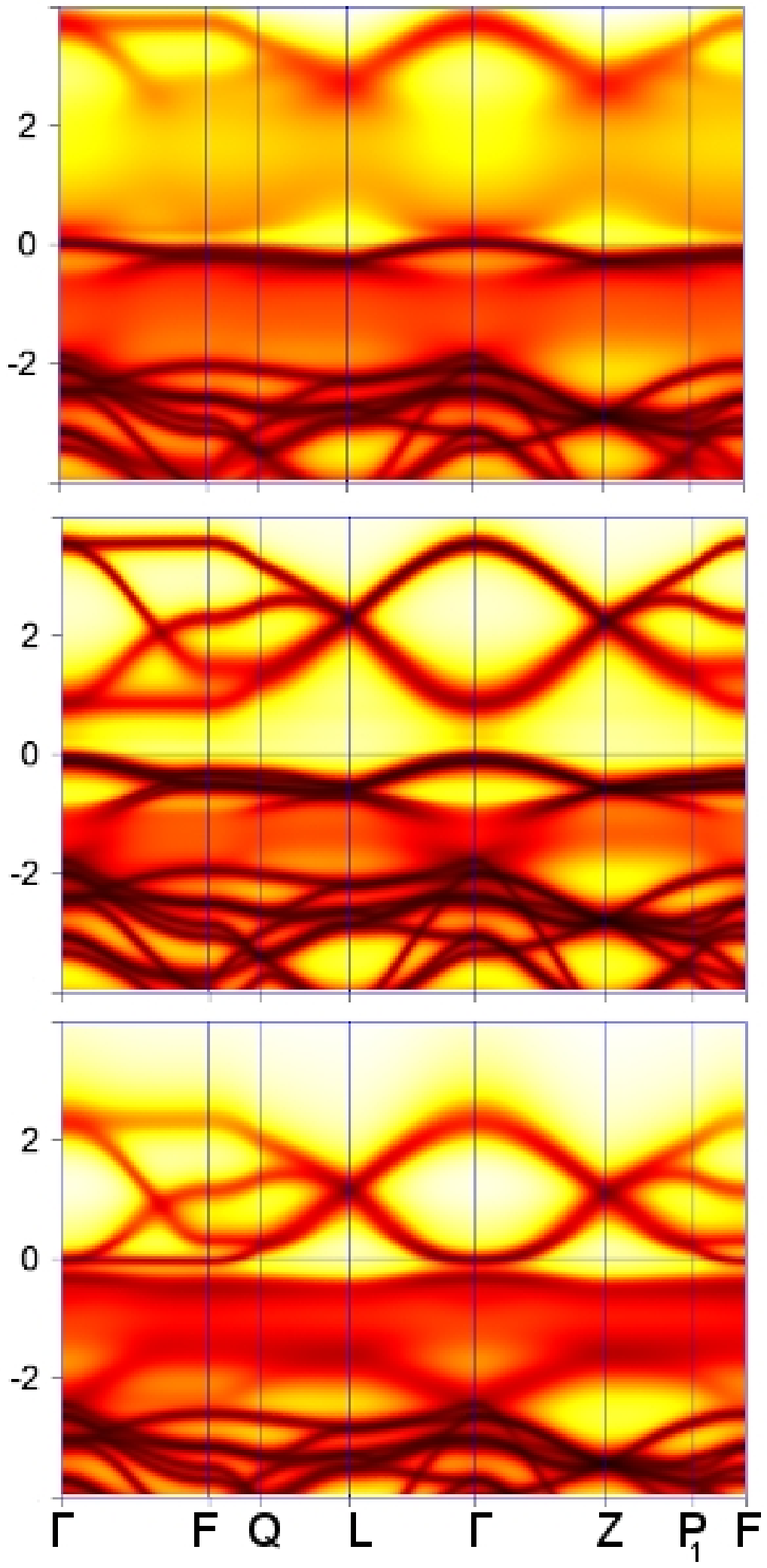}
  \end{minipage}
\caption{\label{fig:spec} (color online) Left panel: orbitally resolved spectral density: Co $e_g$ (black),
Co $t_{2g}$ (red), O $p$ (shaded blue), $p-\pi$ (lighter blue) for various dopings.
Right panel: the corresponding k-resolved spectral functions along
the high symmetry lines in the Brillouin zone depicted as color plots.}
\end{figure}

The one-particle spectral functions decomposed into their
momentum and orbital contributions are shown
in Fig.~\ref{fig:spec}. The stoichiometric system exhibits sharp
bands with well defined dispersion. 
While the electron doping introduces also some correlation effects we focus on the hole doping
which is relevant for La$_{1-x}$Sr$_x$CoO$_3$. 
At x=0.2 the chemical potential is pinned close to the top of the $t_{2g}$ band.
The $e_g$ spectral weight is transferred from the coherent
$e_g-p$ anti-bonding band into incoherent structures around the chemical potential and
fills the gap. 
The k-dependent spectrum clearly shows the spectral weight transfer cannot be described
as a simple band shift due to an orbitally dependent potential and thus has to be attributed
to dynamical correlation effects.
With the number of holes growing the chemical potential shifts
deeper into the $t_{2g}$ band, which exhibits increasing mass renormalization.
At higher doping levels some $e_g$ spectral weight builds up at the chemical 
potential. These are strongly damped excitations with a lifetime 
15~times shorter than that of $t_{2g}$ excitations ($-\operatorname{Im}\Sigma_{e_g}(0)\approx 0.6$~eV,
$-\operatorname{Im}\Sigma_{t_{2g}}(0)\approx 0.04$~eV).
For the full evolution of the hole doped spectra up to $x=0.3$ see SM.

The spectral weight redistribution in the conduction band can be observed
by the X-ray absorption spectroscopy (XAS). Small core-hole effects 
at the oxygen K-edge spectrum allow a direct comparison of the XAS data with the O~$p$ spectral 
function, which follows the $d$ spectrum due to $p$--$d$ hybridization.~\cite{kurmaev08}
The overall shape of the calculated O~$p$ spectrum is unaffected by doping
and compares well with the photoemission data of Saitoh {\it et al.}~\cite{saitoh97} 
including the positions of the non-bonding ($\sim -3$~eV) and the $\sigma$-bonding peaks ($\sim -5.5$~eV).
\begin{figure}
  \centering
  \begin{minipage}{0.43\columnwidth}
   \centering
   \includegraphics[height=0.75\columnwidth,angle=270,clip]{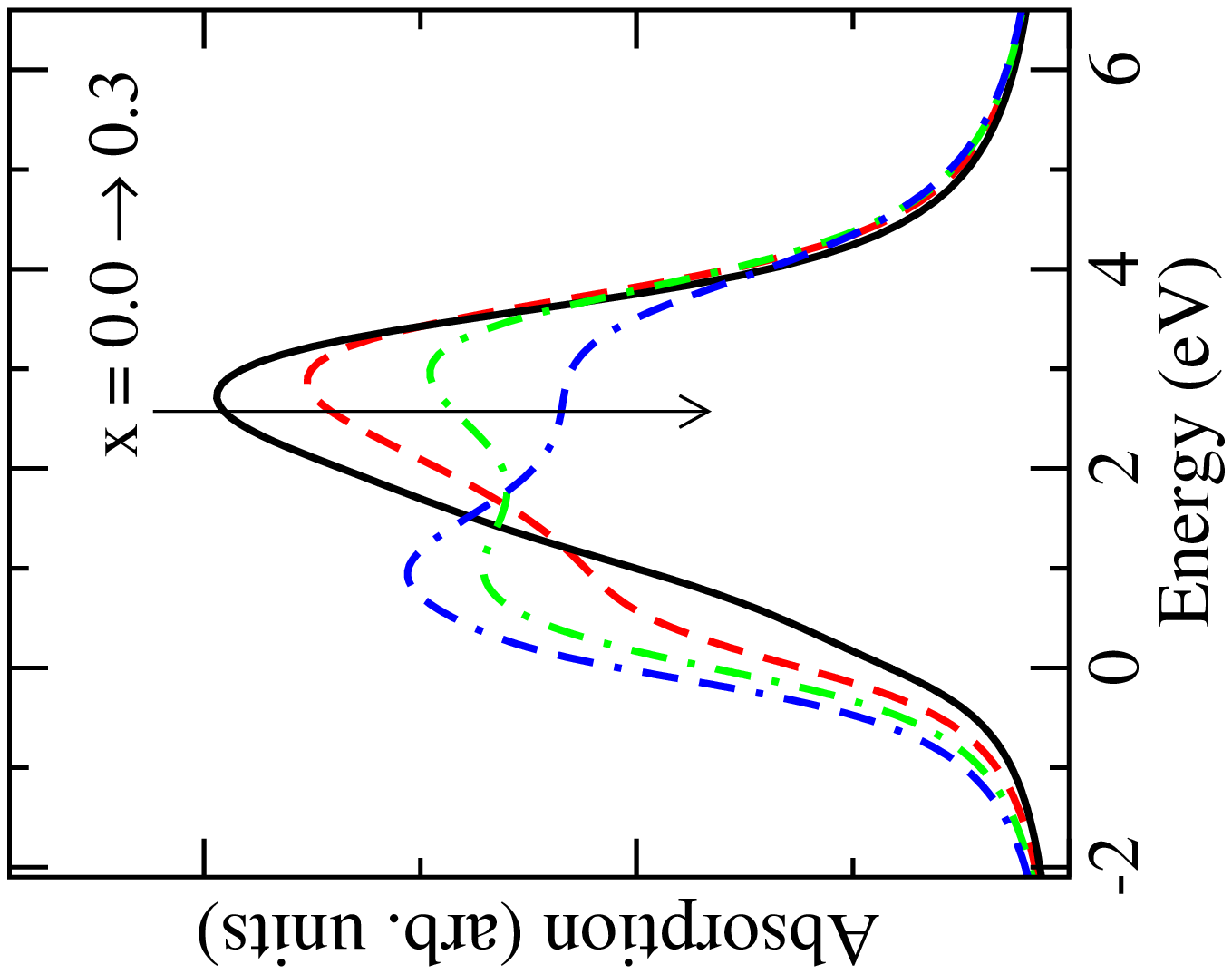}
  \end{minipage}
  \begin{minipage}{0.55\columnwidth}
  \includegraphics[width=\columnwidth,clip]{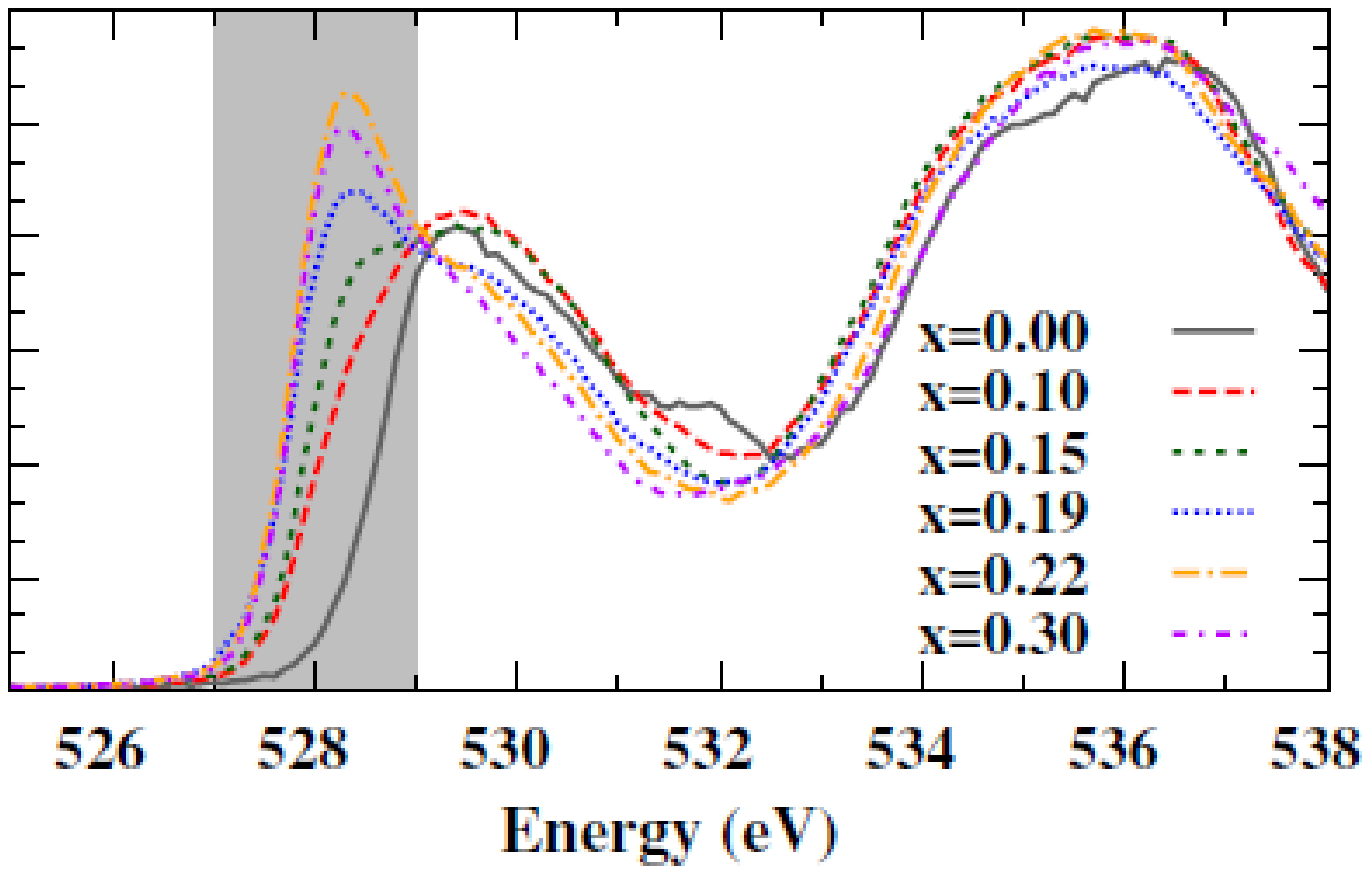}
  \end{minipage}
\caption{\label{fig:p} Left: the unoccupied O $p$ spectra (broadened with Lorentzian)
 for various hole dopings.
 Right: The experimental XAS spectra reproduced from Ref.~\onlinecite{medling12}.}
\end{figure}
In Fig.~\ref{fig:p} we compare the positive energy part of the O $p$
spectra to the recent experimental data of Medling {\it et al.}~\cite{medling12}
While the LSDA theory of Medling {\it et al.} can account for the increase
of the low-energy peak while keeping the high-energy peak unchanged upon doping, 
our LDA+DMFT data reproduce the spectral weight transfer between
the two peaks accompanied by appearance of an isosbestic point.~\cite{greger13}
The increase of the low-energy peak arises from the depopulation of the
$t_{2g}$ states as well as from the transfer of the $e_g$ spectral weight, which gradually
depletes the high-energy peak.

\begin{figure}
  \centering
  \begin{minipage}{0.45\columnwidth}
   \centering
   \includegraphics[height=0.95\columnwidth,angle=270,clip]{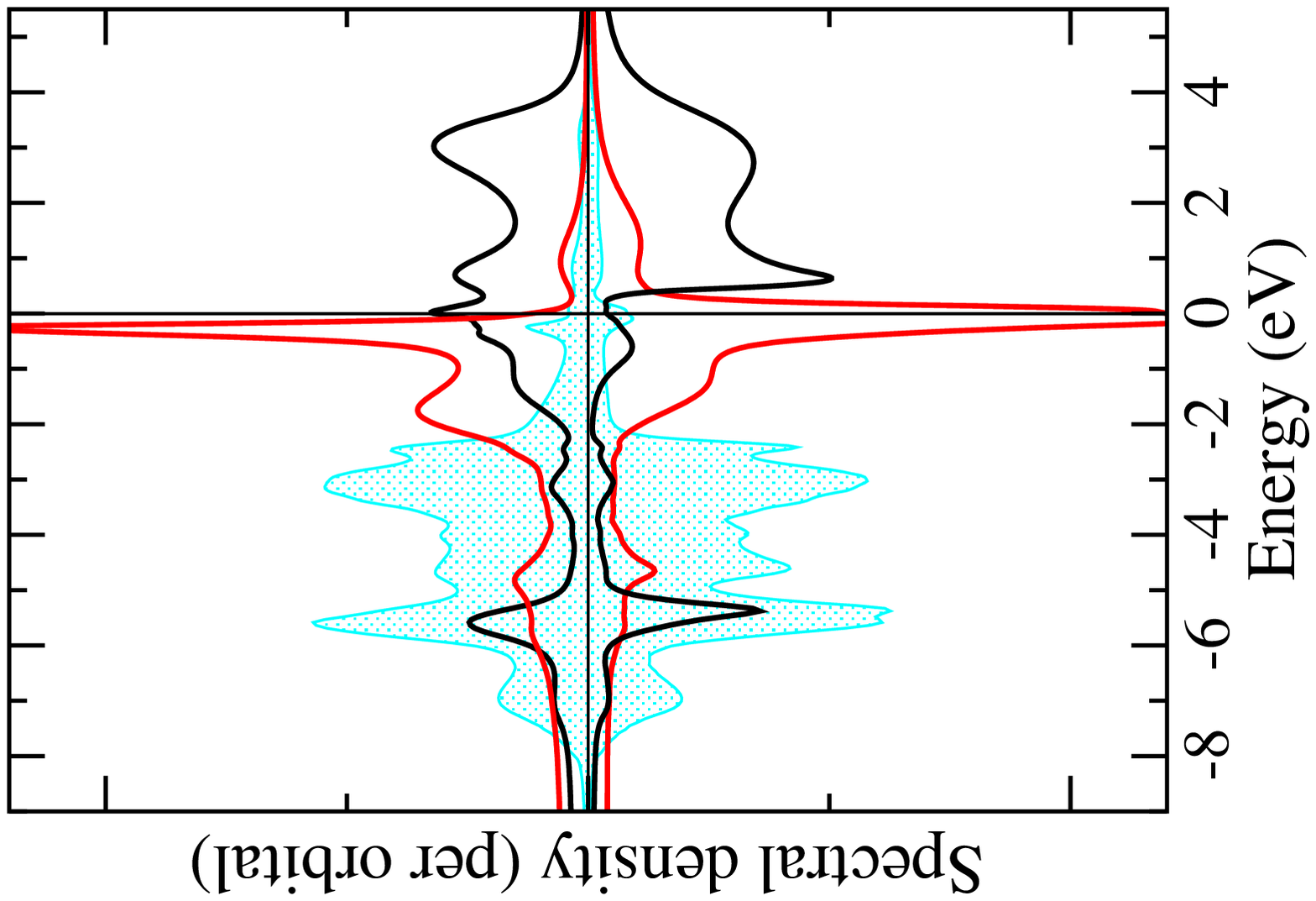}
  \end{minipage}
  \begin{minipage}{0.45\columnwidth}
  \includegraphics[width=0.95\columnwidth,clip]{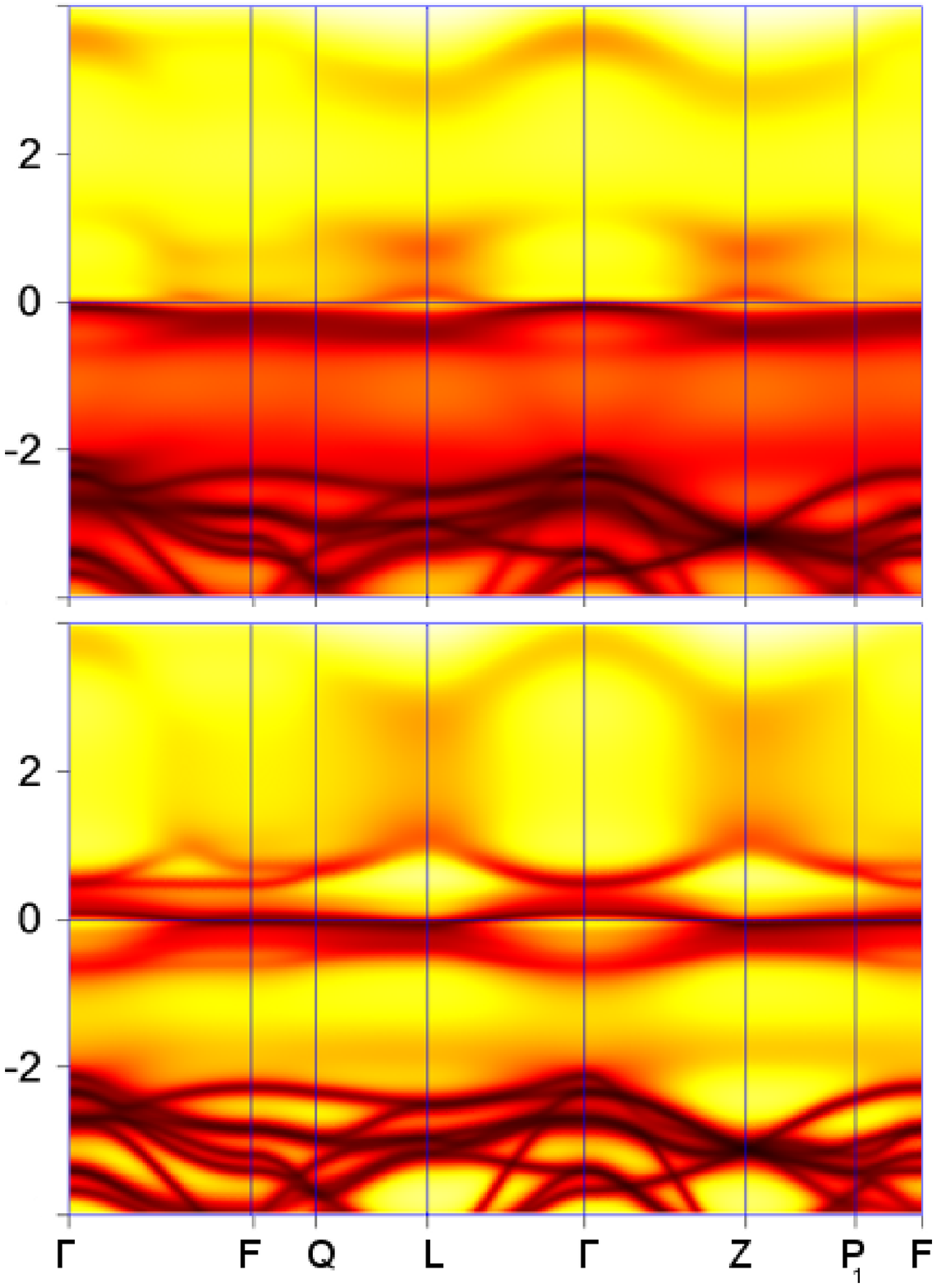}
  \end{minipage}
\caption{\label{fig:spec_fm} (color online) Left panel: orbitally resolved spectral density: Co $e_g$ (black),
Co $t_{2g}$ (red), O $p$ (shaded blue), $p-\pi$ (lighter blue) 
for the ferromagnetic solution obtained at $T$=580~K and hole doping x=0.3. (Minority spin densities
are multiplied with $-1$). Right panel: The corresponding k-resolved spectra 
for the majority (top) and minority (bottom) polarization.}
\end{figure}

Next, we discuss magnetic order.
At the hole doping $x=0.3$ there is a stable FM solution for $T$=580~K. The
ordered moment of 1.2~$\mu_B$ per formula unit is close to saturation as was 
confirmed by calculation at lower temperature.
This is slightly below the experimental value of saturated magnetization.~\cite{kriener04,onose06}
The spectral functions are shown in Fig.~\ref{fig:spec_fm}. Similar 
to the spectrum of SrCoO$_3$~\cite{kunes12},
the majority $t_{2g}$ and minority $e_g$ states are absent from the chemical potential, while
the minority $t_{2g}$ states form a well defined band, which is the main conduction channel.
The majority $e_g$ states form a strongly damped band which crosses the chemical potential.
The spin-averaged spectra exhibit only minor difference to the PM spectra, with the $e_g$ lifetime
by a factor of two larger in the FM phase (see SM for details).
The atomic state correlation
matrices show no sizable redistribution of the multiplet weights, in particular
no enhancement of the IS block, due to the spin polarization. 
This shows that inter-site FM correlations $\langle S_i S_j\rangle$, which are completely
absent in the PM solution, do not stabilize the IS state.

La$_{1-x}$Sr$_x$CoO$_3$ is commonly compared to perovskite manganites and
the double-exchange model is invoked with the partially populated $e_g$ band
providing the conduction channel. In this setting the $d^6$ IS state
is a natural building block maximizing the kinetic energy gain.
Our results provide a picture which differs in two important aspects:
(i) the HS state is identified as the dominant contributor to the CW susceptibility,
(ii) $t_{2g}$ QPs are found at the Fermi level while the $e_g$ excitations
are strongly damped. In the rest of the paper we discuss our results
in the light of the known experimental data.

We studied only homogeneous solutions with full crystal periodicity. The results
are, therefore, not relevant for the spin-state polaron regime at very low doping.~\cite{louca03,podlesnyak08}
It is plausible that while IS states participate in the polaron formation
they are not important at higher doping levels. The polaron is stable
if the kinetic energy gain from delocalizing an $e_g$ electron
over the polaron volume outweighs the cost of exciting the participating atoms
into the IS state. While the latter term grows linearly with
the size of the polaron, the kinetic energy gain is largest for small polarons
and quickly saturates as the polaron grows. Therefore only small polarons
are stable and a substantial change of magnetic properties around $x=0.05$ \cite{podlesnyak11}
suggests that quite different microscopic physics is involved.
Other experimental facts pointing against the IS scenario in the metallic phase ($x>0.2$)
are the lack of substantial magnetoresistance (found in manganites),~\cite{paras01}
no Griffiths phase,~\cite{he07} and observation of positive curvature of
inverse susceptibility above $T_c$.~\cite{he07} The last observation points
to increase of the paramagnetic moment with temperature, while in the
IS scenario one would expect decrease of IS atoms in magnetically disordered
phase. 

Instead of the double-exchange mechanism stabilizing the spin-state
polarons, which relies on the spin order on nano-scale, we
attribute the observed population of the HS state to 
the hybridization contribution to the crystal field on Co.
Introducing a hole on a given Co site is similar to 
a substitution with a more electronegative element. It results
in more covalent bonding with its O neighbors which in turn weakens the covalent 
bonding between those and the next-neighbor Co atoms 
promoting the HS states on them.~\cite{kyomen03,knizek12} 
Since the holes are mobile this picture should be viewed
only as a simplified static analog of a complex dynamical 
equilibrium.

Another, perhaps the most important, result 
is the observation of much stronger damping of $e_g$ excitations
than their $t_{2g}$ counterparts. The orbital decoupling~\cite{demedici09,demedici11}
by strong $J$ together with closer-to-half filling of $e_g$ orbitals 
suggests an explanation. However, unlike the system in Ref.~\onlinecite{demedici11}
our $e_g$ bands do not exhibit an enhanced QP renormalization
but rather a pronounced departure from Fermi liquid behavior. We also do not
observe a suppression of the inter- or intra-orbital fluctuations (see SM).
Therefore we offer a different explanation based on
the observation that the effect is still present in the FM solution, where
the occupied majority $t_{2g}$ and empty minority $e_g$ orbitals do not play an active role,
and that a similar effect (including the behavior of orbital occupancies)
is observed~\cite{kunes13} in a much simpler two-band model of Ref.~\onlinecite{kunes11} when hole doped.
The situation can be viewed as two partially filled asymmetric bands of
spinless fermions. The second Born approximation for the self-energy
in the imaginary time
\begin{equation}
\label{eq:born}
\Sigma_{\sigma}(\tau)=U^2G_{\sigma}(\tau)G_{-\sigma}(\tau)G_{-\sigma}(\beta-\tau), 
\end{equation}
where $\sigma=\pm1$ indexes the two bands, shows that the ratio of the self-energies
at $\beta/2$ is the inverse ratio of the Green functions. This approximately
means that the band with lower density at the Fermi level has more strongly damped
QPs. In other words, the slow $t_{2g}$ holes act as `static' scattering centers 
for the fast $e_g$ electrons, while the $t_{2g}$ holes perceive the $e_g$ electrons as `homogeneous gas'.

As a direct consequence of the $e_g$ damping there is only an inverted-U-shaped QP band 
($t_{2g}$) crossing the chemical potential while its U-shaped $e_g$ counterpart is missing, 
Figs.~\ref{fig:spec},~\ref{fig:spec_fm}. The prediction of this characteristic spectral feature can be
studied with angle-resolved photoemission spectroscopy. Other consequences
of the $e_g$ damping, which involve transport coefficients, are less direct
allowing only speculative comments at the moment.
The present solution is consistent with the absence of colossal magnetoresistance. The anomalous Hall effect
observed in FM metallic phase \cite{onose06} may be related to the
spin-orbit effects in the $t_{2g}$ QP band crossing the Fermi level.
Finally, the contribution of both $t_{2g}$ and $e_g$ channels with
different QP damping, of which the $e_g$ one is more sensitive to the magnetic
order, should be considered when interpreting the $T$-dependence of thermopower.~\cite{berggold05}

In summary, we have used the dynamical mean-field theory
to study hole doping of lanthanum cobaltite.
LaCoO$_3$ is shown to be an example of material in which the electronic 
correlations are hidden to one-particle spectroscopy at low temperature, 
but become manifest upon doping. 
Depletion of electrons from the system leads to growing population 
of the nominally empty $e_g$ orbitals in striking contrast to 
the behavior of weakly correlated materials.
The LDA+DMFT results provide a good description of the spectral and magnetic
properties of the homogeneous metallic phase for $x>0.2$ where
the IS state is shown not to play an important role and the CW susceptibility
is attributed to the HS state and fluctuations around it. We found coherent $t_{2g}$
bands crossing the Fermi level, while the $e_g$ excitations are strongly damped
and appear at the Fermi level only at dopings $x>0.2$. This leads to the conclusion
that the double-exchange model with the $e_g$ band as the conduction channel
is not appropriate for doped cobaltites.

We thank P. Nov\'ak, Z. Jir\'ak, and J. Hejtm\'anek for numerous discussions.
This work was supported by the Deutsche Forschungsgemeinschaft through FOR1346.
V.\,K. was partly supported by European Social Fund, grant No. CZ.1.07/2.3.00/30.0005.


\end{document}